\documentclass[a4paper,12pt]{article}
\usepackage{epsfig}
\usepackage{graphicx}
\usepackage{subfigure}

\newcommand{\p}{\varphi}
\newcommand{\pr}{\partial}

\def\ie{{\it i.e.,}}

\begin{document}

\renewcommand{\thefootnote}{\fnsymbol{footnote}}
\title{Skyrme Model with Different Mass Terms}
\author{
Bernard Piette\thanks{email: b.m.a.g.piette@durham.ac.uk}\,
and
Wojtek J. Zakrzewski\thanks{email: w.j.zakrzewski@durham.ac.uk}
\\
Department of Mathematical Sciences, University of Durham, \\
Durham DH1 3LE, UK\\
}
\date{\today}

\maketitle

\setlength{\footnotesep}{0.5\footnotesep}
\begin{abstract}
We consider a one parameter family of mass terms for the Skyrme model
that disfavours shell-like configurations for multi-baryon classical solutions.
We argue that a model with such mass terms can provide a better description of 
nuclei as shell like configurations  
are now less stable than in the traditional massive Skyrme model. 

\end{abstract}

Initially proposed by Skyrme as a fundamental model for baryons \cite{Skyrme}, 
the Skyrme model has subsequently been shown \cite{Witten} to be the low energy 
limit of QCD in the $1/N_c$ expansion. 
In dimensionless units, the Hamiltonian for the Skyrme model is given by 
\begin{eqnarray}
E&=&\frac{1}{12\pi^2}\int_{R^3}\left\{-\frac{1}{2}\mbox{Tr}\left(\pr_iU\,
U^{-1}\right)^2+\frac{1}{16}\mbox{Tr}\left[\pr_iU\,
U^{-1},\;\pr_j U\, U^{-1}\right]^2\right\}d^3\vec{x}\nonumber\\
&&
+\frac{1}{12\pi^2}\int_{R^3} \,m\sp2\, \mbox{Tr}(1-U)\,d^3\vec{x},
\label{skham}
\end{eqnarray}
where $m$ denotes the pion mass.

Recently the model has been used also in the description of nuclei (as a
semiclassical model describing their properties) \cite{BMS}.
The major problem with all the applications
of the model is associated with the fact that 
the energy densities of its classical solutions \cite{BS}  have 
shell like configurations with a hole in the middle, even for a 
relatively large number of baryons, unless one takes for the pion mass 
$m$ a value several times 
larger than the physical one. To avoid this problem it has been suggested that 
one should consider the pion mass in the Lagrangian as a free parameter that 
must be fitted to the experimental data, thus somehow justifying the use of 
unphysical values of $m$.

Recently, we argued \cite{KPZ} that the mass term in the Skyrme model 
is not unique
and that there exist a large family of such terms that have the correct 
asymptotic behaviour to describe pion fields. 
In general, the mass term,  \ie  the last term in (\ref{skham}), which we call 
the potential, can be written as
\begin{equation} 
V=\frac{1}{12\pi^2}Am_\pi\sp2\,\int_{R^3} \, \mbox{Tr}
  \left[1-\int_{-\infty}\sp{\infty}\,g(p) \, U\sp{p} \,dp\,
\right]\,d^3\vec{x},
\label{genmass}
\end{equation}
where
\begin{equation} 
\int_{-\infty}\sp{\infty} \,g(p)  \, dp\,=\, 1\,\quad \mbox{and}\quad
\int_{-\infty}\sp{\infty} \,g(p)p\sp2
 \, dp\,=\, 1
\end{equation}
and 
\begin{equation} 
A\sp{-1}\,=\,\int_{-\infty}\sp{\infty} \,g(p)p\sp2
  \, dp.
\end{equation}

In \cite{KPZ} we looked in detail at the special case when $g(x)=\delta(x-p)$
for integer values of $p$ and showed that when $p$ is even and non-zero (note 
$p=0$ is different), the classical 
solutions are very much like in the massless case and form hollow shells with a 
vanishing energy density at their centre. When $p$ is odd all
shell-like configurations have a non-vanishing energy density at the centre
which, if $m$ and $B$ are large enough,  disfavours shell-like configurations. 
The energy density at the centre of the configuration is the largest  
when $p=1$ which is the conventional mass term.

Of all the mass terms that we analysed in \cite{KPZ} the only term which 
disfavours shell-like configurations more than the traditional mass term 
is the special case $p=0$, which, however,  corresponds to a 
non-analytical $U$ potential term.

In this paper, we have decided to look at a one parameter family of mass 
terms which are
a linear combination of three of the mass terms studied in 
\cite{KPZ} chosen in such a way that the shell like configurations are 
disfavoured.

To achieve this,  we need a mass term that asymptotically
leads to the usual expression for the pion fields in the linear limit, 
$m_\p |\vec{\pi}|^2$, but which is larger inside the shell in the baryon sector.
We can obtain a family of such terms by taking (\ref{genmass}) with
\begin{equation} 
g(p) = \delta(p-1)+D( \delta(p-2)- \delta(p-3)).
\end{equation}
Notice that in this case we have 
\begin{equation} 
A^{-1} = 1 +D(4 -9) = 1-5\,D
\end{equation} 
and
\begin{equation}
\int_{-\infty}\sp{\infty} \,g(p)  \, dp\,=\, 1+D(1-1) \,=\,1.
\end{equation} 
So our potential term reduces to the explicit expression
\begin{equation} 
V = \frac{1}{12\pi^2} \frac{m_\pi^2}{1-5D} Tr(1-U -D (U^2-U^3)).
\label{massD}
\end{equation}

Computing classical solutions of the Skyrme model when the baryon charge is 
larger than one is very difficult, but Houghton et al. \cite{HMS} have 
presented
the so called ``rational map ansatz'' which makes it possible to compute
good approximations to shell-like solutions of the Skyrme model.
The ansatz involves taking the fields $U$ as
\begin{equation}
U = exp(i f(r) (2 P - 1)),
\label{RM}
\end{equation}
where $f(r)$ is a radial profile function and $P$ is a particular projector 
which depends only
on the angular coordinates $\theta$ and $\varphi$.
Defining $z= e^{i\varphi}\tan\frac{\theta}{2}$ we have 
$P=\frac{v v^\dagger}{|v|^2}$ where
$v= (1,R(z))$ is a two component holomorphic complex vector, 
$\frac{\partial R}{\partial \bar{z}} = 0$. Moreover, $R(z)$ is a rational
function of $z$ and the degree of the rational map corresponds to the baryon
number of the configuration. 
Both the radial profile and the rational map can be determined by inserting
the ansatz (\ref{RM}) into (\ref{skham}) and minimising the obtained 
expression:
\begin{eqnarray}
H &=& {1\over 3 \pi} \int \Big[
   f_r^2 +2 B {\sin^2 f \over r^2}(1+f_r^2) 
    + {\cal I} {\sin^4 f \over r^4}   \nonumber\\
   && + 2{m^2\over 1 - 5D}(1-\cos(f)-D(\cos(2f)-\cos(3f))) \Big]   r^2\,dr, 
\label{RMham}
\end{eqnarray}
where 
\begin{equation}
 {\cal I}=\frac{1}{4\pi}\int \bigg(
\frac{1+\vert \xi\vert^2}{1+\vert R\vert^2}
\bigg\vert\frac{dR}{d\xi}\bigg\vert\bigg)^4 \frac{2i \  d\xi  d\bar \xi
}{(1+\vert \xi\vert^2)^2}\,.
\label{I}
\end{equation}
The integral (\ref{I}) depends only on the rational map $R$ and so it must be 
minimised 
separately from the radial profile. This was done in \cite{HMS} and \cite{BS1}.

Note that choosing a more general mass term like (\ref{genmass})
does not change the angular dependance of the configuration within the 
rational map ansatz. It only changes the equation for the profile function 
$f(r)$.

Before we attempt to minimise (\ref{RMham}) we must check that the
potential is positive definite.
It easy to show this but only when $D$ is 
in the range $[0, 0.2[$. 
Indeed, the roots of $(1-\cos(f) -D (\cos(2f)-\cos(3f)))$
are given by $f=0$ and $D=1/(4\cos^2(f)+2\cos(f)-1)$. 
If we exclude the point $f=0$, the energy density can only vanish  
if $D \le 0$ or if $D \ge 1/5$. Notice that the case $D=0$ corresponds to the 
standard mass term.

To compute the low energy configurations of our model that  approximate its 
solutions we have solved the following equation
for the radial profile $f(r)$
\begin{eqnarray}
f_{rr}(1&+&2 B{\sin^2 f \over r^2}) + 2 {f_r \over r}
+ B {\sin (2f) \over r^2} (f_r^2 -1) -I{\sin(2f) \sin^2 f\over r^4} 
\nonumber\\
&-&{m^2 \over 1-5D} (\sin (f)+D (2\sin(2f)-3\sin(3f)))  = 0,
\label{eqprofmod}
\end{eqnarray}
taking from \cite{BS1} the value for ${\cal I}$.

\begin{figure}[ht]
\unitlength1cm \hfil
\begin{picture}(16,16)
 \epsfxsize=8cm \put(0,8){\epsffile{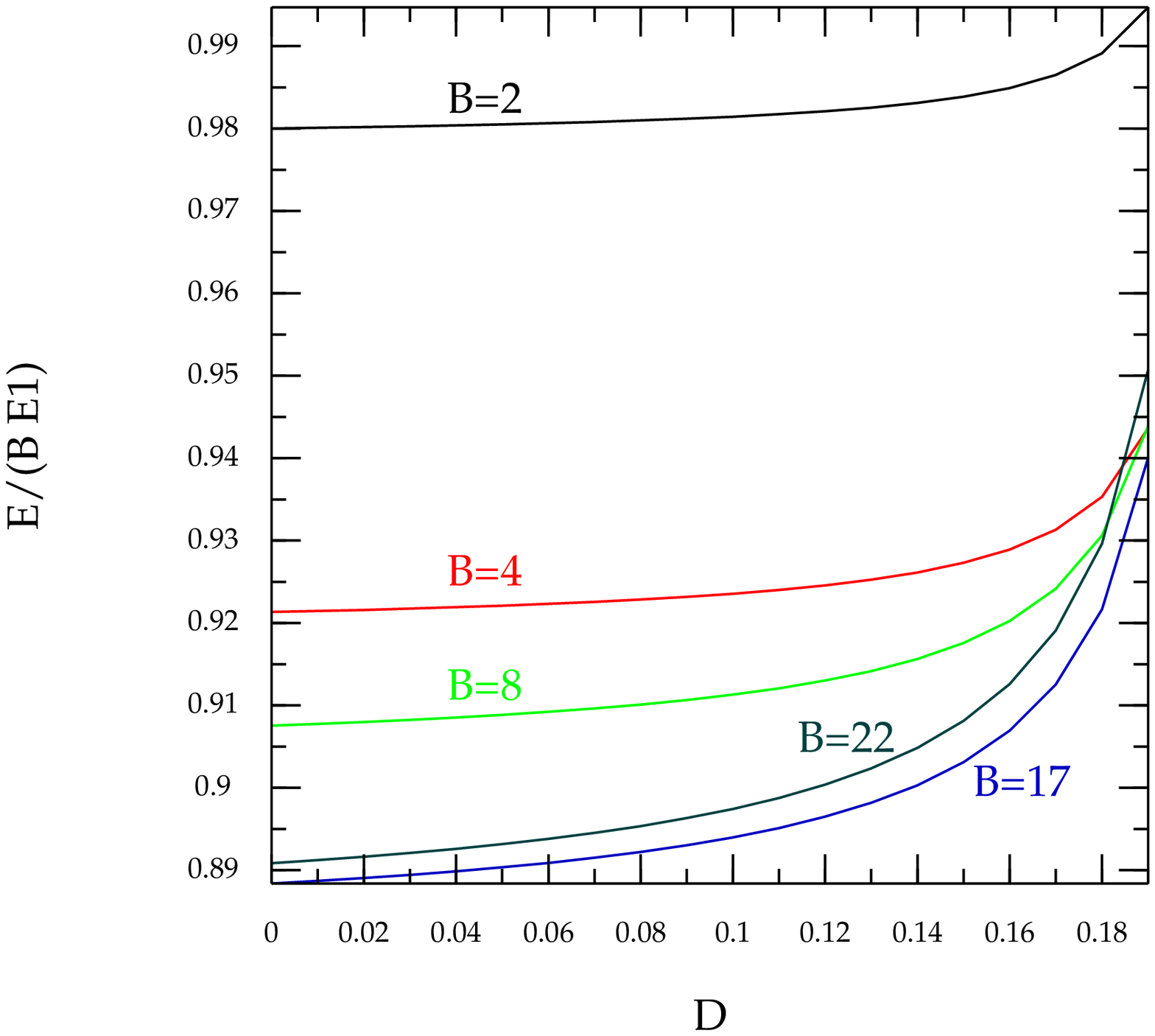}}
 \epsfxsize=8cm \put(8,8){\epsffile{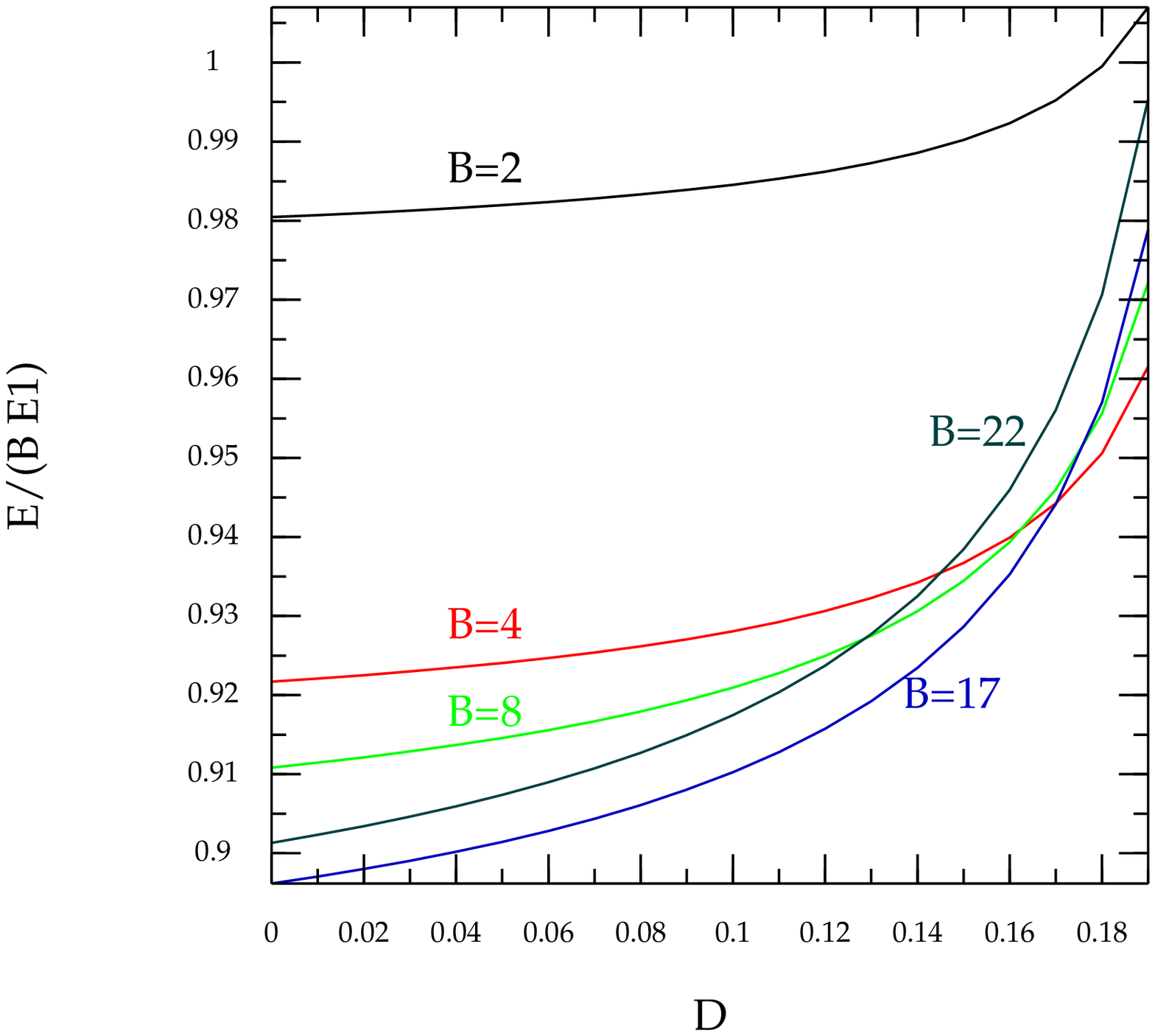}}
 \epsfxsize=8cm \put(0,0){\epsffile{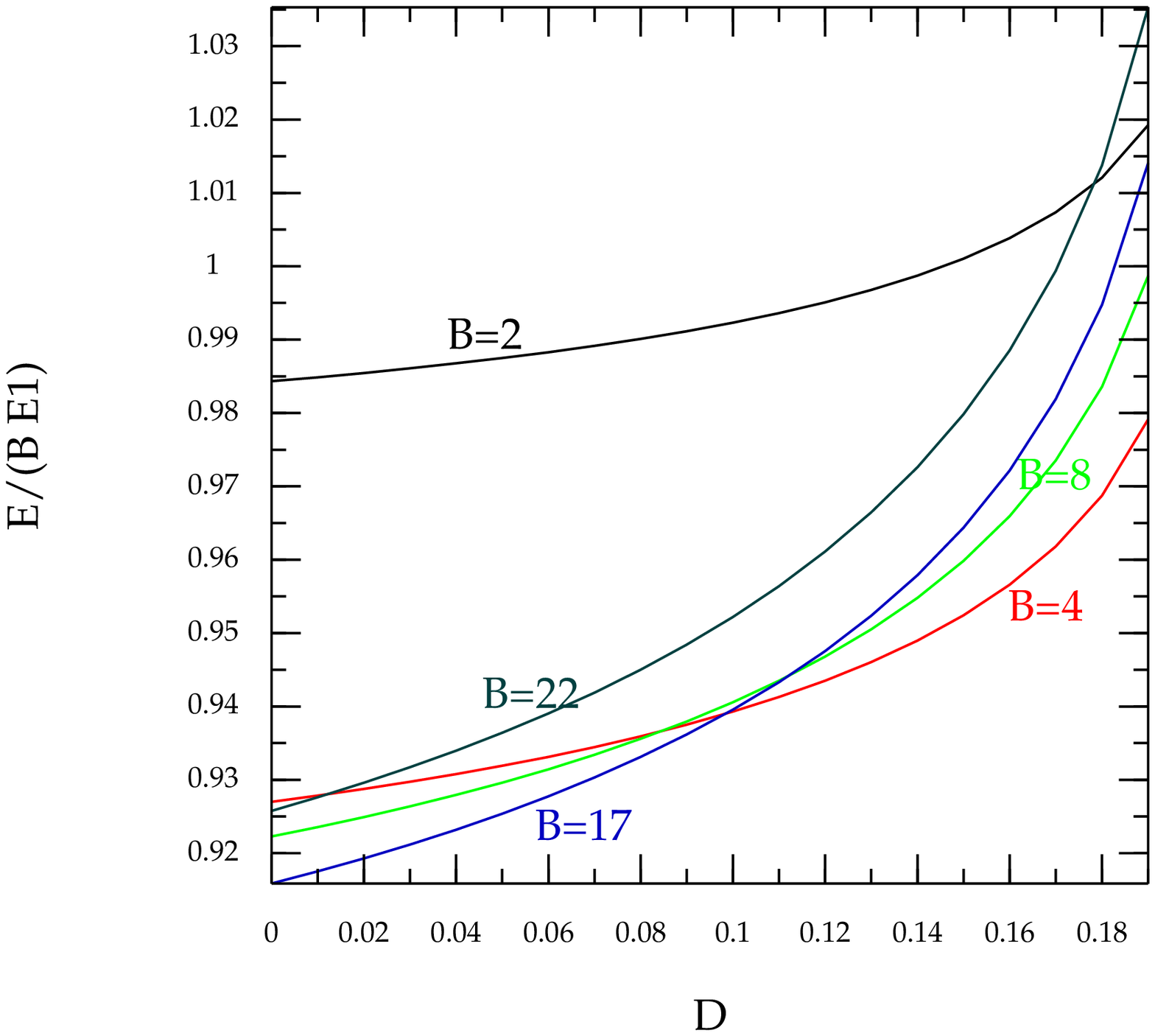}}
 \epsfxsize=8cm \put(8,0){\epsffile{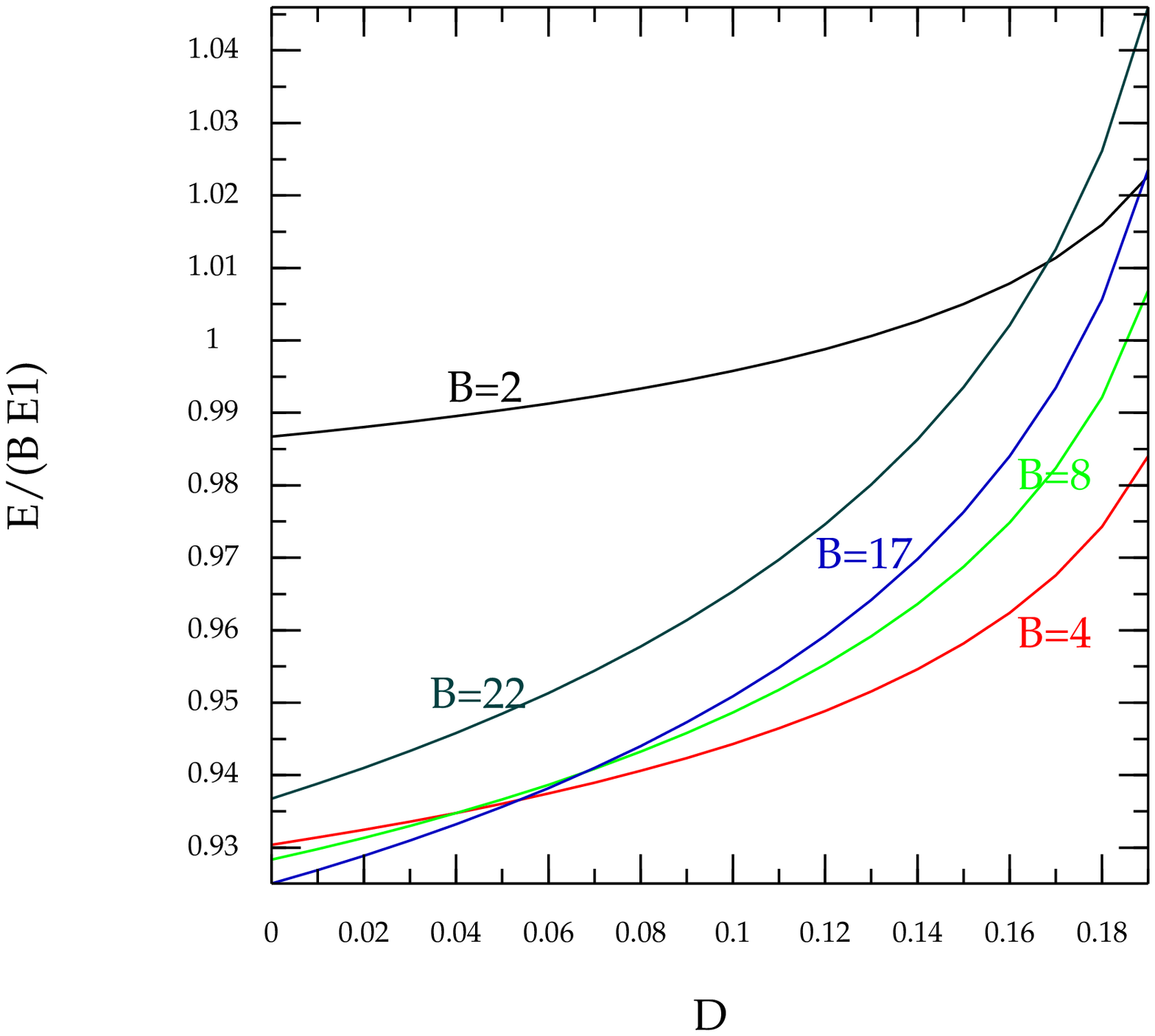}}
\put(4.0,8){a}
\put(12.0,8){b}
\put(4.0,0){c}
\put(12.0,0){d}
\end{picture}
\caption{\label{fig1}
Relative baryon energies : $E(B)/BE(1)$ as a function of $D$
a) $m=0.2$, b) $m=0.4$, c) $m=0.8$; d) $m=1$.
}
\end{figure}

\begin{figure}[ht]
\unitlength1cm \hfil
\begin{picture}(16,16)
 \epsfxsize=8cm \put(0,8){\epsffile{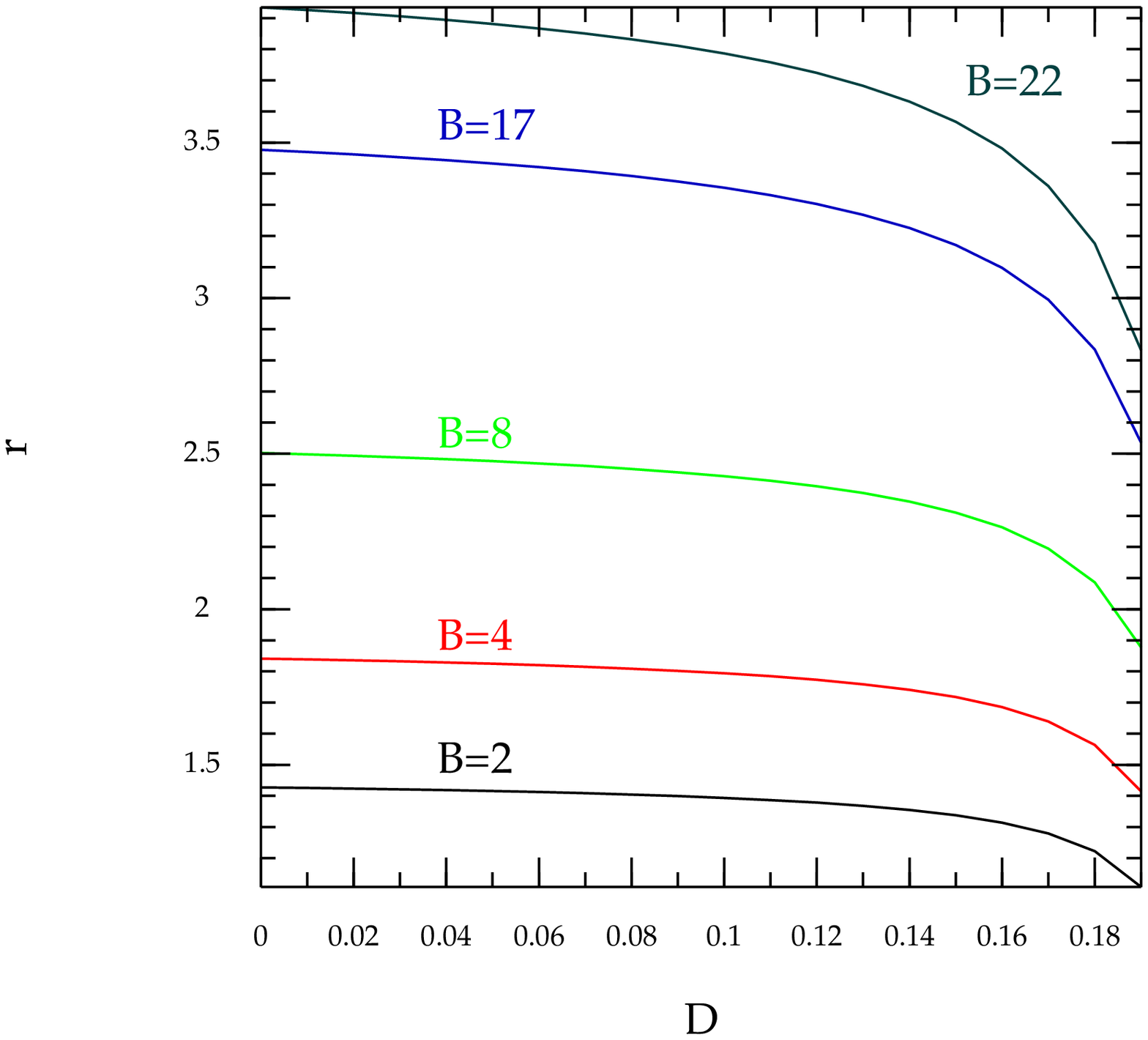}}
 \epsfxsize=8cm \put(8,8){\epsffile{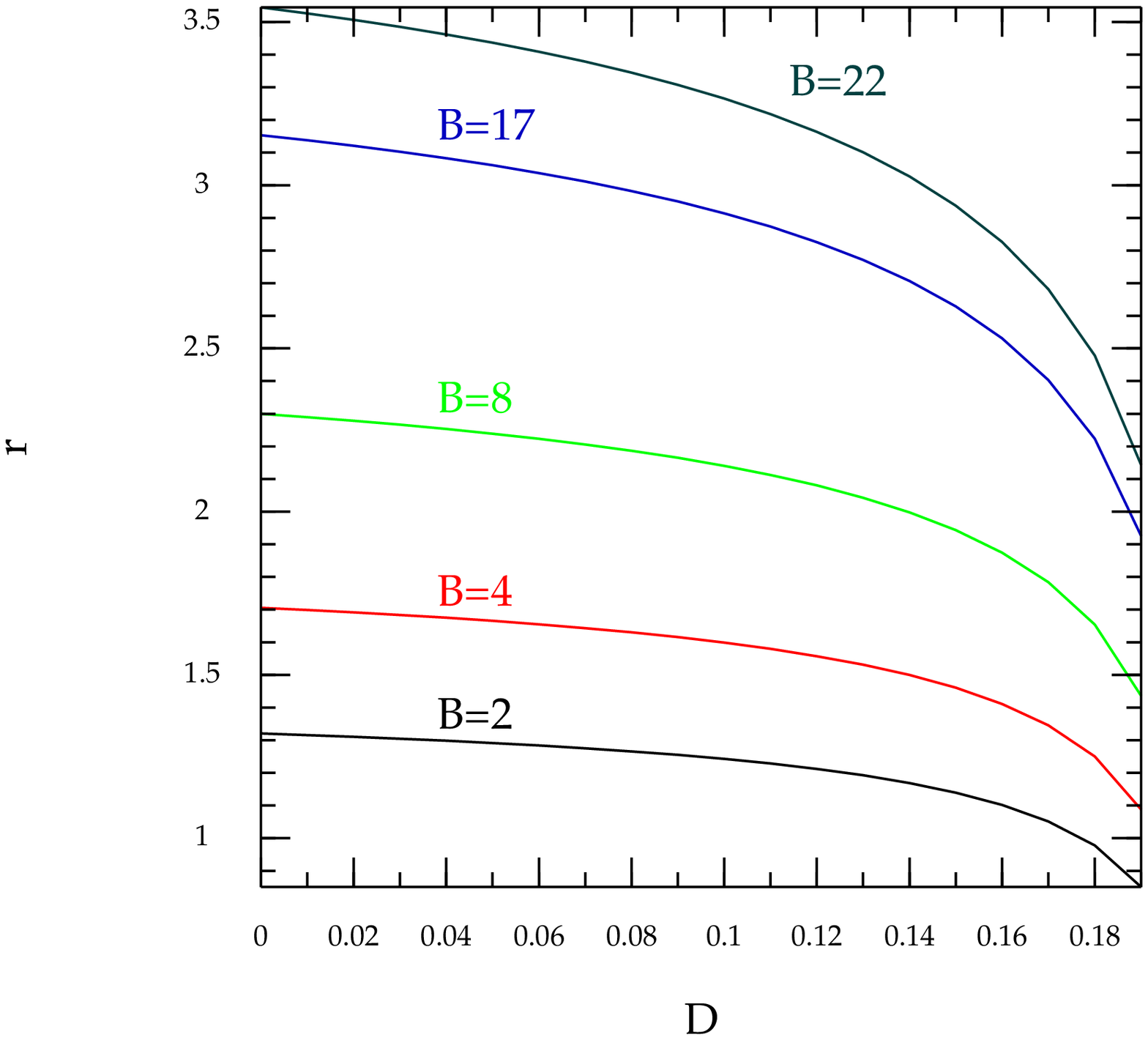}}
 \epsfxsize=8cm \put(0,0){\epsffile{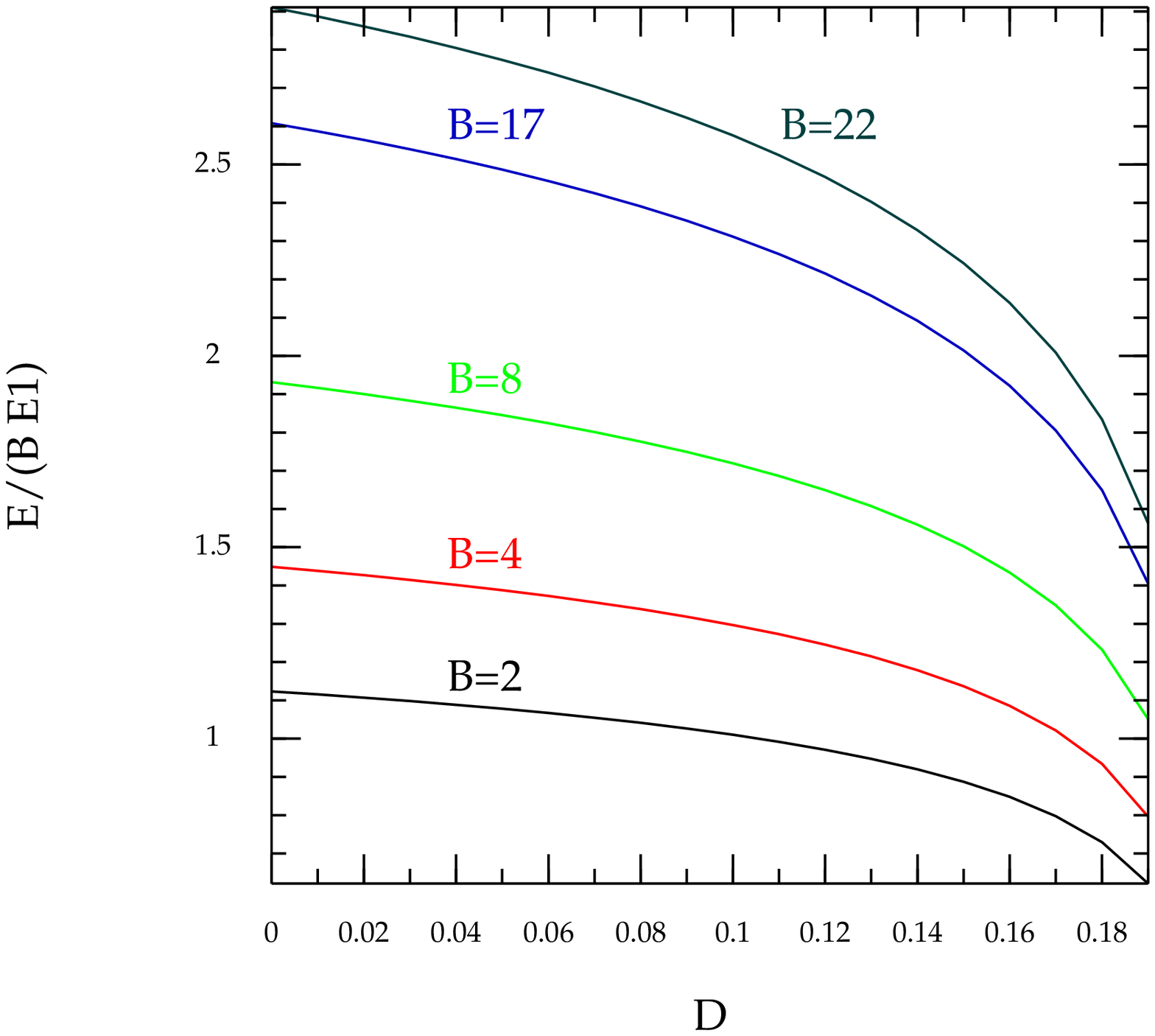}}
 \epsfxsize=8cm \put(8,0){\epsffile{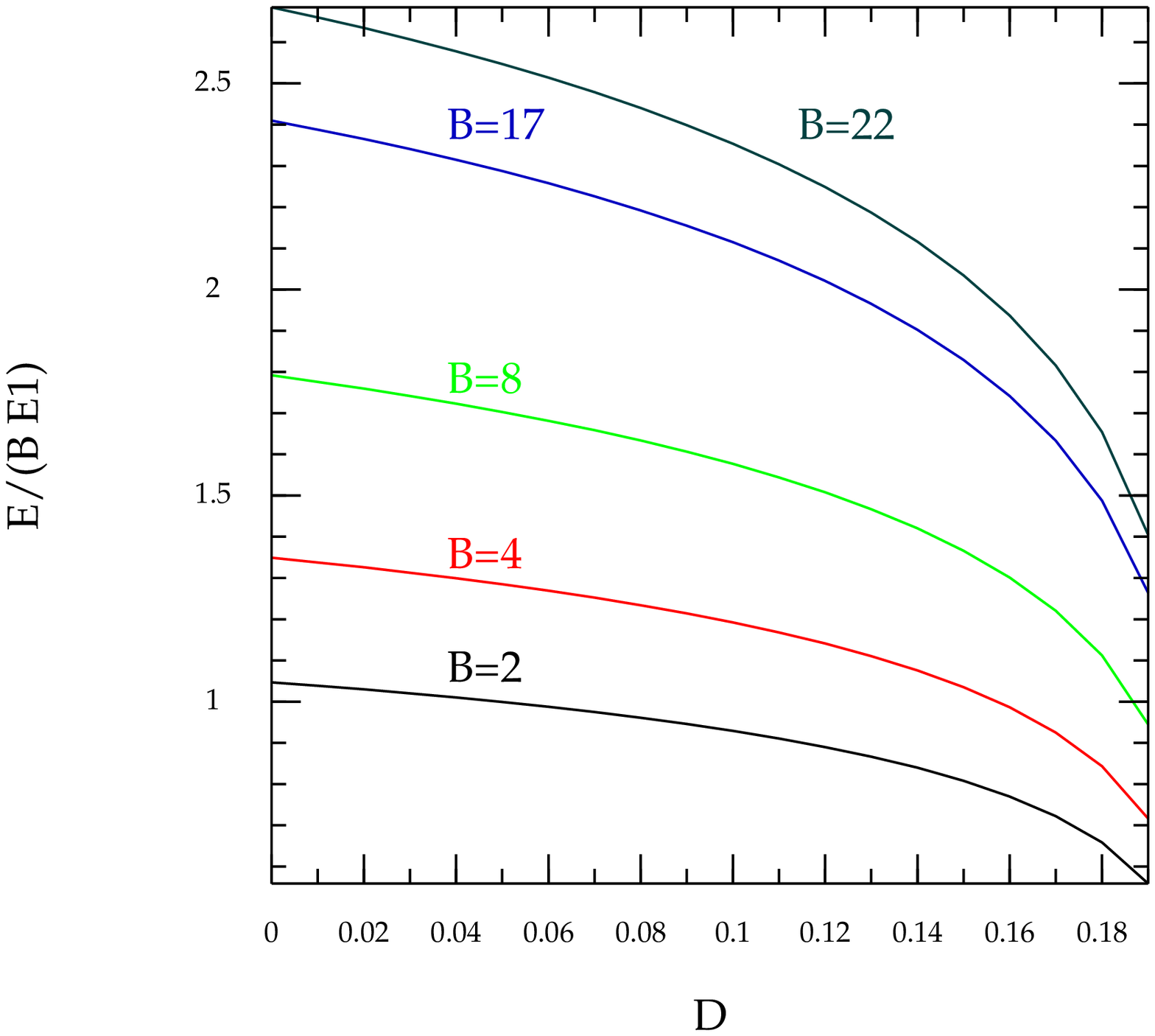}}
\put(4.0,8){a}
\put(12.0,8){b}
\put(4.0,0){c}
\put(12.0,0){d}
\end{picture}
\caption{\label{fig2}
Shell radius : 
$ r=\int_0^\infty \rho E(\rho) \rho^2 d\rho/\int_0^\infty E(\rho) \rho^2 d\rho$ 
as a function of $D$
a) $m=0.2$, b) $m=0.4$, c) $m=0.8$; d) $m=1$.
}
\end{figure}

The results are shown in figure \ref{fig1} where we have plotted, as a 
function of $D$, the energy of 
several solutions divided by $B$ times the energy of the corresponding $B=1$
solution. This relative energy describes how bound the skyrmions of the
corresponding solution are. 

We see clearly that, for a fixed value of the pion mass $m$, the relative 
energy per baryon increases with $D$. This shows that the potential we have 
chosen dislikes shell-like configurations more than the standard mass term.
We also notice that for a relatively small value of the mass, $m=0.4$ 
(Figure \ref{fig1}b ), the curves for the 
relative energies of the $B=17$ and $B=22$
configurations both cross the curves for the $B=4$ and $B=8$ cases
showing that these shell configurations could decay into smaller baryon 
configurations. Of course we expect the existence 
of solutions which are not shell like and which thus cannot be approximated by 
the rational map ansatz and we expect that the shell-like configurations 
would decay into these solutions instead of into the multiple shells.
Even the configuration $B=8$ is unstable with respect
to the decay into two $B=4$ configurations when $D>0.16$. As the value of $m$ 
increases, we see that the instability of the shell-like configurations 
increases.

In figure  \ref{fig2} we present the radius, 
$ r=\int_0^\infty \rho E(\rho) \rho^2 d\rho/\int_0^\infty E(\rho) \rho^2 d\rho$, 
of the energy denisity of the configuration, as a function of $D$,
 for different values of the mass $m$.
We see that in each case, the radius decreases with $D$. This shows that the 
energy density at the center of the Skyrmion configuration increases with $D$,
forcing the minimal energy configuration to have a smaller radius.

We have shown that if one takes (\ref{massD}) as the mass term for the Skyrme 
model, one obtains a model that disfavours shell-like configurations much more 
than the model with the traditional mass term,
 which corresponds to (\ref{massD}) with $D=0$. 
This comes from the fact that the energy density at the centre of the shell
increases with $D$ and, as a result, it is more favourable for the 
configuration 
to have a non-hollow shell-like shape (the fields do not take values close 
to $U=-1$ over a larger region than in the traditional model). Such non-shell 
like configurations have been constructed by Battye et al. \cite{BMS}\cite{BS}
for the traditional mass term, but to obtain such configurations, 
Battye et al. had to take a value of $m$ 6 or 7 times larger than the 
physical value of the pion mass.

To provide a semi-classical description of nuclei the Skyrme model must have 
solutions that are not hollow at their centres.
 While Battye et al. have shown that one can 
do this by taking a large value of the pion mass, we have shown that one can 
achieve the same effect with a more physical value of the pion mass by taking 
a different mass term for the Skyrme model {\it i.e.} (\ref{massD}). 

\vskip 5mm
{\Large \bf Acknowledgement}

We would like to thanks P. Sutcliffe for drawing our attention to the fact that 
the case $p=0$ in \cite{KPZ} corresponds to a non-analytic potential.
We would also like to thank V. Kopeliovich for useful discussions on this 
topic.


\begin{thebibliography}{99}
\bibitem{Skyrme} T.H.R. Skyrme, 
Proc. Roy. Soc. Lond., {\bf A260}, 127-138 (1961).
\bibitem{Witten} E. Witten, Nucl. Phys. {\bf B223}, 422 (1983), 
       E. Witten, Nucl. Phys. {\bf B223}, 433 (1983)
\bibitem{BMS} R.A. Battye, N. Manton and P.M. Sutcliffe, 
Proc. R. Soc. {\bf 463}, 261 (2007)
\bibitem{BS} R.A. Battye and P.M. Sutcliffe, 
Nucl. Phys. {\bf B705}  384 (2005)
\bibitem{KPZ} V.B. Kopeliovich, B. Piette, and W.J. Zakrzewski, 
Phys. Rev. D {\bf 73}, 014006 (2006) 
\bibitem{HMS} C.J. Houghton, N.S. Manton and P.M. Sutcliffe, 
Nucl. Phys. B {\bf 510}, 507 (1998).
\bibitem{BS1} R.A. Battye and P.M. Sutcliffe, 
Rev. Math. Phys. {\bf 14}, 29 (2002)
\end{thebibliography}
\end{document}